\documentclass[sigconf,screen]{acmart}
\renewcommand\footnotetextcopyrightpermission[1]{} 
\AtBeginDocument{%
  }

\copyrightyear{2025}
\acmYear{2025}
\setcopyright{cc}
\setcctype{by}
\acmConference[CIKM '25]{Proceedings of the 34th ACM International Conference on Information and Knowledge Management}{November 10--14, 2025}{Seoul, Korea}
\acmBooktitle{Proceedings of the 34th ACM International Conference on Information and Knowledge Management (CIKM '25), November 10--14, 2025, Seoul, Korea}
\acmDOI{}
\acmISBN{}

\begin{document}

\title[Multi-modal Adaptive Mixture of Experts for Cold-start Recommendation]{Multi-modal Adaptive Mixture of Experts for \\ Cold-start Recommendation}

\author{Van-Khang Nguyen}
\orcid{0009-0000-3854-9515}
\affiliation{%
  \institution{VNU University of Engineering and Technology}
  \city{Hanoi}
  \country{VietNam}
}
\email{21020768@vnu.edu.vn}

\author{Duc-Hoang Pham}
\orcid{0009-0007-6847-384X}
\affiliation{%
  \institution{VNU University of Engineering and Technology}
  \city{Hanoi}
  \country{VietNam}
}
\email{22021200@vnu.edu.vn}

\author{Huy-Son Nguyen}
\orcid{0009-0006-4616-0976}
\affiliation{%
  \institution{Delft University of Technology}
  \city{Delft}
  \country{The Netherlands}
}
\email{H.S.Nguyen@tudelft.nl}

\author{Cam-Van Thi Nguyen}
\orcid{0009-0001-9675-2105}
\affiliation{%
  \institution{VNU University of Engineering and Technology}
  \city{Hanoi}
  \country{VietNam}
}
\email{vanntc@vnu.edu.vn}

\author{Hoang-Quynh Le\authornotemark{}}
\orcid{0000-0002-1778-0600}
\authornote{Corresponding author.}
\affiliation{%
  \institution{VNU University of Engineering and Technology}
  \city{Hanoi}
  \country{VietNam}
}
\email{lhquynh@vnu.edu.vn}

\author{Duc-Trong Le}
\orcid{0000-0003-4621-8956}
\affiliation{%
  \institution{VNU University of Engineering and Technology}
  \city{Hanoi}
  \country{VietNam}
}
\email{trongld@vnu.edu.vn}


\begin{abstract}
Recommendation systems have faced significant challenges in cold-start scenarios, where new items with a limited history of interaction need to be effectively recommended to users. Though multimodal data (e.g.,\textit{ images, text, audio, etc}.) offer rich information to address this issue, existing approaches often employ simplistic integration methods such as concatenation, average pooling, or fixed weighting schemes, which fail to capture the complex relationships between modalities. Our study proposes a novel Mixture of Experts (MoE) framework for multimodal cold-start recommendation, named MAMEX, which dynamically leverages latent representation from different modalities. MAMEX utilizes modality-specific expert networks and introduces a learnable gating mechanism that adaptively weights the contribution of each modality based on its content characteristics. This approach enables MAMEX to emphasize the most informative modalities for each item while maintaining robustness when certain modalities are less relevant or missing. Extensive experiments on benchmark datasets show that MAMEX outperforms state-of-the-art methods in cold-start scenarios, with superior accuracy and adaptability. For reproducibility, the code has been made available on Github.\footnote{\url{https://github.com/L2R-UET/MAMEX}}.
\end{abstract}





\maketitle

\section{Introduction}
Recommendation systems have become indispensable in modern digital ecosystems, enabling personalized content delivery in e-commerce, streaming services, and social media~\cite{liu2024multimodal,nguyen2024bundle,malitesta2025formalizing}. However, these systems often face the cold-start problem~\cite{volkovs2017dropoutnet, du2020learn, bai2024multimodality,bui2025personalized}, particularly in item cold-start scenarios, where recommendations must be made for new items with little or no interaction data. In these cases, traditional collaborative filtering methods usually do not work well because they depend on user-item interaction history to make accurate suggestions~\cite{bui2025personalized,bai2024multimodality}.
To address this issue, recent research \cite{bai2024multimodality, liu2024alignrec, wei2021contrastive, chen2022generative} has explored the integration of multimodal data, such as product images, textual descriptions, and audio features, which offer complementary information that can enrich item representations and improve the precision of the recommendation, especially in item cold-start scenarios. For example, the image of a fashion product can convey a visual style, while its textual description provides semantic attributes such as material or brand. Using such diverse modalities, models can make more informed predictions even when behavioral data are sparse.
Despite this potential, effectively fusing multimodal information remains a nontrivial task. Many existing methods \cite{bai2023gorec, liu2024alignrec, zhou2023contrastive,zhou2023enhancing,nguyen2025ramen} rely on straightforward fusion techniques, such as concatenation or averaging, which treat each modality equally and independently. These approaches often overlook the inherent differences in modality characteristics and fail to capture complex cross-modal relationships effectively. Moreover, they lack adaptability in assigning importance to different modalities across varying items, which is crucial in heterogeneous data environments and even more so in item cold-start scenarios. These limitations motivate the need for a more adaptive and content-aware fusion strategy.
In this work, we propose \textbf{MAMEX} (\textit{Multimodal Adaptive Mixture of Experts}), a novel recommendation framework that addresses the limitations of conventional multimodal fusion by leveraging the Mixture of Experts (MoE) paradigm. Our model introduces a multi-stage expert architecture that dynamically adapts to the content structure of each item, allowing it to selectively emphasize the most informative modalities.

\vspace{-2mm}
\section{Related Work} \label{related-work}

Cold-start recommendation is challenging due to the scarcity of user-item interaction ~\cite{xu2024cmclrec, bui2025personalized}. To mitigate this, multimodal methods incorporate heterogeneous data such as images and text~\cite{ganhor2024multimodal, malitesta2025formalizing,nguyen2025ramen,zhou2023enhancing}. Yet, conventional fusion approaches, like simple concatenation~\cite{nguyen2023hhmc,zhou2023enhancing} or averaging~\cite{nguyen2025ramen}, often fail to adequately capture the complex relationships among different modalities~\cite{bai2023gorec,liu2024multimodal}. Although attention mechanisms improve fusion, they struggle with variable quality or missing modalities, limiting effectiveness.
The Mixture of Experts (MoE) framework, which adaptively combines specialized networks via a gating function~\cite{mu2025comprehensive}, has shown promise in multi-task learning~\cite{zhang2024m3oe} and sequential recommendation~\cite{bian2023multi} but remains underexplored for multimodal cold-start recommendation~\cite{bui2025personalized, zhang2025hierarchical, bian2023multi}. Some methods apply MoE to handle each modality independently, demonstrating its potential in modeling sequential and modality-specific signals. However, these works typically treat modalities in isolation and lack effective cross-modal interaction modeling. Thus, the integration of MoE to jointly fuse multiple modalities, especially under cold start conditions, remains underexplored.

Our work addresses this gap with a dual-level MoE framework combining modality-specific experts and cross-modal interaction via a dynamic gating mechanism. Furthermore, we introduce a balance regularization term to prevent modality collapse and enhance robustness in diverse cold-start scenarios.


\section{Methodology} \label{method}
\begin{figure*}[htp]
    \centering
    \hspace{0.05\textwidth}
    \includegraphics[width=0.9\textwidth]{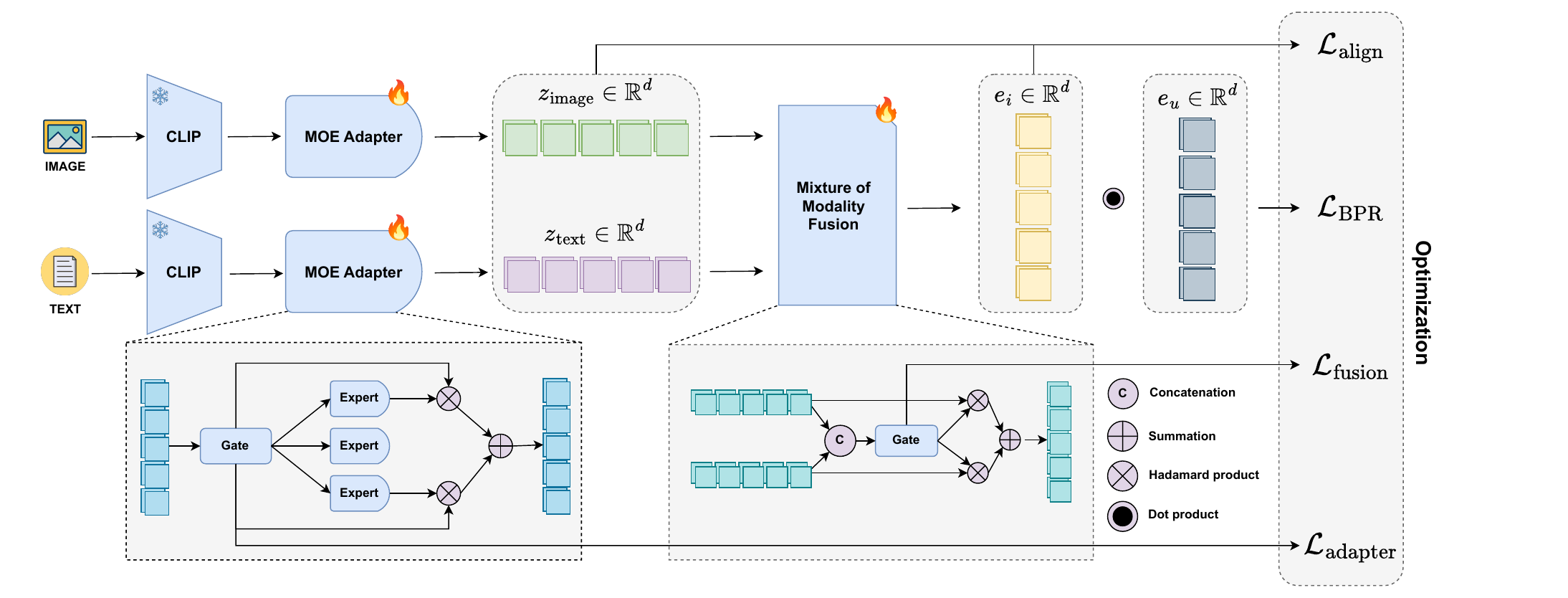}
    \caption{The overview architecture of our proposed framework MAMEX.}
    \label{fig:overall_model}
\end{figure*}

The MAMEX architecture, as shown in Fig.~\ref{fig:overall_model}, consists of $2$ key modules: a Modality Extraction Module that processes and aligns individual modality features; and a Modality Fusion Module that adaptively combines these features leveraging a gating mechanism.

\subsection{Preliminaries}
Let $\mathcal{U},\mathcal{I}$ denotes the sets of users and items, respectively. Each item $i \in \mathcal{I}$ has modality-specific raw features $\{x_i^1, x_i^2, \ldots, x_i^{|\mathcal{M}|}\}$, each from a distinct modality. Within $\mathcal{I}$, we identify a subset $\mathcal{I'} \subset \mathcal{I}$ comprising newly added items with little or no user feedback. User-item interactions are represented by a sparse matrix $\mathcal{X} \in \mathbb{R}^{|\mathcal{U}| \times |\mathcal{I}|}$, where each entry $x_{u,i}$ denotes the presence or intensity of interaction between user $u$ and item $i$.
The objective is to learn a multimodal recommendation function $g: \mathcal{U} \times \mathcal{I'} \rightarrow \mathbb{R}$ that estimates the relevance score between users and cold-start items:
\begin{equation}
y_{u,i'} = g(u, i', \{x_{i'}^1, x_{i'}^2, \ldots, x_{i'}^{|\mathcal{M}|}\}, \mathcal{X})
\end{equation}
where $y_{u,i'}$ is the predicted relevance between $u$ and $i' \in \mathcal{I'}$. 

\subsection{Modality Extraction Module}
\label{modality_extraction_module}
We process each modality with specialized extractors and refine them via a MoE to align and enhance features.
\subsubsection{Feature Extraction}
Each modality $m$ is extracted initial feature representations through modality-specific pre-trained models:
\begin{equation}
h_m = E_m(x_m)
\end{equation}
where $E_m$ represents the feature extractor for modality $m$ (e.g., CLIP~\cite{radford2021learning} for image and text, wav2vec~\cite{baevski2020wav2vec} for audio), and $x_m$ denotes the raw for that modality. After extraction, the input data $x_m$ is transformed into a modality representation $h_m \in \mathbb R^{{d}}$.

\subsubsection{Modality-Specific Adaptation}
To better adapt the extracted features for recommendation, we introduce a modality-specific MoE layer with $K$ expert networks and a gating mechanism that dynamically weights experts based on the input:
\begin{equation}
z_m = \sum_{k \in \mathcal{T}} \bigg( \hat{g}_m^k(h_m)E_m^k(h_m) \bigg)
\end{equation}

where $E_m^k(h_m) \in \mathbb{R}^d$ denotes the $k$-th expert network, implemented as a linear transformation. The sparse gating weights $\hat{g}_m^k$ select only the top-$k$ experts with indices in $\mathcal T$, where $\mathcal T$ represents the indices of experts with the highest gating value.

To mitigate the challenge of expert underutilization in MoE framework, we propose a load balancing loss that encourages more uniform data allocation across experts:

\begin{equation}
\mathcal{L}_{\text{adapter}} = \sum_{m \in \mathcal M} D_{\text{KL}}\left(\frac{1}{N}\sum_{i=1}^{N} g_m^k(h_m^i) \bigg\| \frac{1}{K}\mathbf{1}\right)
\end{equation}
where $D_{\text{KL}}(P \parallel Q)$ is the Kullback-Leibler divergence, which quantifies the discrepancy between two probability distributions. The first term inside the divergence denotes the empirical average of the gating function outputs across a batch of input samples, effectively capturing the actual routing probability distribution over $K$ experts. The target $\frac{1}{K}\mathbf{1}$ is a uniform distribution over experts, ensuring equal expert utilization.


\subsection{Mixture of Modality Fusion}
\label{moe_fusion}
The second component of MAMEX combines the representations of aligned modality adaptively through a dynamic fusion mechanism.

\subsubsection{Adaptive Fusion Mechanism}
We form unified item representations by weighted-summing embeddings from all available modalities. For image and text, weights are computed via a sparse softmax gating function \(G(\cdot)\) over their concatenation:
\begin{align}
\alpha &= G\Big(z_{\text{image}} \parallel z_{\text{text}}\Big)
\end{align}
Based on the modality features and their corresponding weights, the final representation of item is calculated as follow: 
\begin{align}
e_i = \sum_{m \in \mathcal{M}} \alpha_m \cdot z_m
\end{align}
where \(\mathcal{M}\) denotes the set of modalities (e.g., \(\{\text{image}, \text{text}\}\)), $z_m\in \mathbb R^d$ are modality-specific embeddings, and \(\alpha_m \in \mathbb R^{m}\) are the corresponding sparse softmax gating weights. 

\subsubsection{Balanced Fusion Regularization}
To mitigate modality collapse, characterized by the predominance of one modality over another, we introduce a balance regularization term:
\begin{align}
\mathcal{L}_{\text{fusion}} = D_{\text{KL}}\Bigg(\frac{1}{N}\sum_{i=1}^{N} \alpha^{(i)} \bigg\| \frac{1}{m}\mathbf{1}\Bigg),
\end{align}
where \(D_{\text{KL}}\) is the Kullback-Leibler divergence, \(\alpha^{(i)}\) represents the fusion weights for the \(i\)-th item, and \(\frac{1}{m}\mathbf{1}\) (with \(m=2\) for the two modalities) denotes a uniform distribution over modalities.

\subsubsection{Modality Alignment Loss}
To ensure that the final representation \(i_{\text{final}}\) captures the semantic traits of each modality, we use a Mean Squared Error-based alignment loss is computed as:
\begin{align}
\mathcal{L}_{\text{align}} = \sum_{m \in \mathcal{M}} ||e_i - z_m||_2^2
\end{align}
 This alignment loss is added to the overall training objective without altering the aggregation formula for modality embeddings.

\subsection{Recommendation Training}
For top-K recommendation, we employ the Bayesian Personalized Ranking loss \cite{rendle2012bpr}. 
Given a user embedding $e_u \in \mathbb R^d$ and the final item embedding $e_i \in \mathbb R^d$, the prediction score is computed using dot product as follows:
\begin{equation}
s_{u,i} = e_u^T e_i
\end{equation}
The Bayesian Personalized Ranking~\cite{rendle2012bpr} loss is subsequently adopted:

\begin{equation}
\mathcal{L}_{\text{BPR}} = \sum_{(u,i,j)\in \mathcal{D}} -\ln \sigma(s_{u,i} - s_{u,j})
\end{equation}
where $(u,i,j)$ is a training triplet with user $u$, positive item $i$, and negative item $j$, and $\sigma$ is the sigmoid function. The total loss is computed by aggregating the individual loss components:
\begin{equation}
\mathcal{L} = \mathcal{L}_{\text{BPR}} + \lambda_1\mathcal{L}_{\text{align}} + \lambda_2\mathcal{L}_{\text{adapter}} + \lambda_3\mathcal{L}_{\text{fusion}} +  \lambda_4||{\Theta}||_2^2
\end{equation}
where \(\lambda_{1-4}\) are hyperparameters weighting the cross-modal alignment, adapter balance, fusion balance, and L2 regularization losses, respectively, and \(\Theta\) represents all trainable parameters.

\section{Results and Discussion} \label{result}
This section presents the experimental results of our approach, including the setup, baseline comparisons, and ablation studies.

\begin{table*}[ht]
\centering
\footnotesize
\caption{Overall performance on Amazon benchmark datasets from across diverse domains such as Baby, Clothing, and Sport. The `bold' numbers present the most outstanding results, while the `\underline{underline}' figures depict the second best performances.}
\label{tab:overall_performance}
\begin{tabular}{lcccc|cccc|cccc}
\toprule
 & \multicolumn{4}{c|}{Amazon Baby} & \multicolumn{4}{c|}{Amazon Clothing} & \multicolumn{4}{c}{Amazon Sport} \\
\cmidrule(lr){2-5} \cmidrule(lr){6-9} \cmidrule(lr){10-13}
Method & Rec@10 & Rec@20 & NDCG@10 & NDCG@20 & Rec@10 & Rec@20 & NDCG@10 & NDCG@20 & Rec@10 & Rec@20 & NDCG@10 & NDCG@20 \\
\midrule
\textbf{MTPR}     & 0.0130 & 0.0246 & 0.0066 & 0.0097 & 0.0215 &0.0377  & 0.0110 & 0.0153 & 0.0146 & 0.0241 & 0.0076 & 0.0102 \\
\textbf{AlignRec} & 0.0224 & 0.0440 & 0.0107 & 0.0166  & 0.0425 & 0.0658 & 0.0234  & 0.0298 & 0.0274 & 0.0465 & 0.0153 & 0.0205 \\
\textbf{MetaEmbed} & 0.0264 & 0.0459 & 0.0132 & 0.0187 & 0.0352 & 0.0602 & 0.0188 & 0.0256 & 0.0349 & 0.0629 & 0.0173 & 0.0250 \\
\textbf{DropoutNet} & 0.0174 & 0.0315 & 0.0083 & 0.0122 & 0.0152 & 0.0277 & 0.0091 &  0.0125 & 0.0171 & 0.0301 & 0.0089 &  0.0125 \\
\textbf{CLCRec}  & 0.0263  & 0.0437 & 0.0136 & 0.0180 &   0.0348 & 0.0481  & 0.0181 & 0.0221 & 0.0271 & 0.0421  &  0.0139 & 0.0177 \\
\textbf{GAR} & 0.0119 & 0.0238 & 0.0057 & 0.0090 & 0.0368 & 0.0629 & 0.0206 & 0.0276 & 0.0360 & 0.0641 & 0.0201 & 0.0278  \\
\textbf{MILK}      & \underline{0.0465} & \underline{0.0730} & \underline{0.0271} & \underline{0.0344} &  \underline{0.0991} & \underline{0.1436} & \underline{0.0571} & \underline{0.0691} & \underline{0.0668} & \underline{0.0998} &  \underline{0.0390} &  \underline{0.0483} \\
\midrule
\textbf{MAMEX}      & \textbf{0.0512} & \textbf{0.0771} & \textbf{0.0292} &  \textbf{0.0363} & \textbf{0.1048} & \textbf{0.1501} & \textbf{0.0606} & \textbf{0.0729}  & \textbf{0.0776} & \textbf{0.1152} & \textbf{0.0470} & \textbf{0.0574}\\


\textcolor{teal}{\textbf{\% Improv.}} & \textcolor{teal}{10.11} & \textcolor{teal}{5.62} & \textcolor{teal}{7.75} & \textcolor{teal}{5.52} & \textcolor{teal}{5.75} & \textcolor{teal}{4.53} & \textcolor{teal}{6.13} & \textcolor{teal}{5.50} & \textcolor{teal}{16.17} & \textcolor{teal}{15.43} & \textcolor{teal}{20.51} & \textcolor{teal}{18.84} \\

\bottomrule
\end{tabular}
\end{table*}

\subsection{Experimental Setup}

\subsubsection{Datasets}
We evaluate our approach on three Amazon Reviews datasets: Baby, Clothing, and Sport \cite{mcauley2015image}. All datasets include product images, textual descriptions, and user reviews. Data are split into $8:1:1$ ratio for training, validation, and test sets and completely remove all interactions for items in the development and test sets to simulate realistic cold-start scenarios.

\subsubsection{Evaluation Metrics}
Inspired by previous works~\cite{bai2024multimodality}, we evaluate performance using Recall@K and NDCG@K for \(K \in \{10, 20\}\).

\subsubsection{Hyperparameter Setup}
Our MoE adapter is tuned over expert numbers $\{4, 6, 8\} $ and top-$k$ routing with $k \in \{2, 3, 4\}$, ensuring both model expressiveness and computational efficiency. We set the learning rate to $0.001$, $\lambda \in \{1, 0.1, 0.01\}$ to mitigate modality collapse, and Adam~\cite{kingma2014adam} for optimization.

\subsubsection{Baseline Methods}
We evaluate our approach against several state-of-the-art methods for cold-start recommendation, such as: {MTPR} \cite{du2020learn}, {AlignRec} \cite{liu2024alignrec}, {CLCRec} \cite{wei2021contrastive}, {GAR} \cite{chen2022generative}, {MILK} \cite{bai2024multimodality}, {DropoutNet} \cite{volkovs2017dropoutnet}, {MetaEmbed} \cite{pan2019warm}.

\subsection{Overall Performance}
Table~\ref{tab:overall_performance} compares MAMEX with state-of-the-art baselines on three Amazon datasets \cite{mcauley2015image}, showing consistent superiority across all metrics. Notably, MAMEX achieves outstanding performance in Recall@10 and NDCG@10: $10.11\%$ and $7.75\%$ on Amazon Baby; $5.75\%$ and $6.13\%$ on Amazon Clothing; $16.17\%$ and $20.51\%$ on Amazon Sport, respectively. These results highlight the effectiveness of our dual-level MoE architecture in capturing modality-specific information while dynamically integrating multimodal signals. 
The gains over all baselines confirm that balance regularization and dynamic gating help prevent modality collapse and improve representation quality, especially in cold-start scenarios.
\subsection{Ablation Studies}

\subsubsection{Impact of Different Components}
To validate the effectiveness of each component in our framework, we conducted ablation studies by selectively removing key components~\footnote{The term `\textit{w/o}' is abbreviation form of `\textit{with out}'.}, such as: \textbf{\textit{w/o} MoE}: Replace the modality-specific MoE layers with standard neural networks; \textbf{\textit{w/o} Alignment}: Remove the modality alignment loss; \textbf{\textit{w/o} MMF}: Replace the adaptive fusion mechanism with simple averaging.


\begin{table}[h]
    \centering
    \caption{Ablation Study Results on Amazon Baby Dataset.}
    \label{tab:ablation}
    \resizebox{\linewidth}{!}{ 
    \begin{tabular}{lcc|cc|cc}
        \toprule
        \textbf{Datasets} & \multicolumn{2}{c|}{\textbf{Baby}} & \multicolumn{2}{c|}{\textbf{Clothing}} & \multicolumn{2}{c}{\textbf{Sports}} \\
        \midrule
        \textbf{Metrics} & \textbf{Rec@20} &\textbf{ NDCG@20} & \textbf{Rec@20} & \textbf{NDCG@20} & \textbf{Rec@20} &\textbf{ NDCG@20} \\
        \midrule
        \textit{w/o} MoE & 0.0746 & 0.0339 & 0.1388 & 0.0647 & 0.0983 & 0.0462 \\
        \textit{w/o }Align & 0.0571 & 0.0260 & 0.1152 & 0.0545 & 	0.0832 & 0.0399 \\
        \textit{w/o} MMF   & \underline{0.0752} & \underline{0.0353} & \underline{0.1429} &  \underline{0.0679} &\underline{0.1084} & \underline{0.0528} \\
        \midrule
        MAMEX & \textbf{0.0771} & \textbf{0.0363} & \textbf{0.1501} & \textbf{0.0729} & \textbf{0.1152} & \textbf{0.0574} \\
        \bottomrule
    \end{tabular}
    } 
\end{table}


As shown in Table~\ref{tab:ablation}, the results clearly indicate that the removal of any key component leads to performance degradation. In particular, omitting the MoE layers or the adaptive fusion mechanism results in significant drops in NDCG@10 (approximately 5.4\% and 4.8\% respectively). Notably, excluding the cross-modal alignment loss leads to the largest performance decline, emphasizing its critical role in bridging the gap between different modalities. These findings confirm that each component contributes complementarily to the overall effectiveness of our model.

\subsubsection{Impact of single modality vs. multi-modalities}

\begin{figure}[h]
    \centering
    \resizebox{0.45\textwidth}{!}{\includegraphics{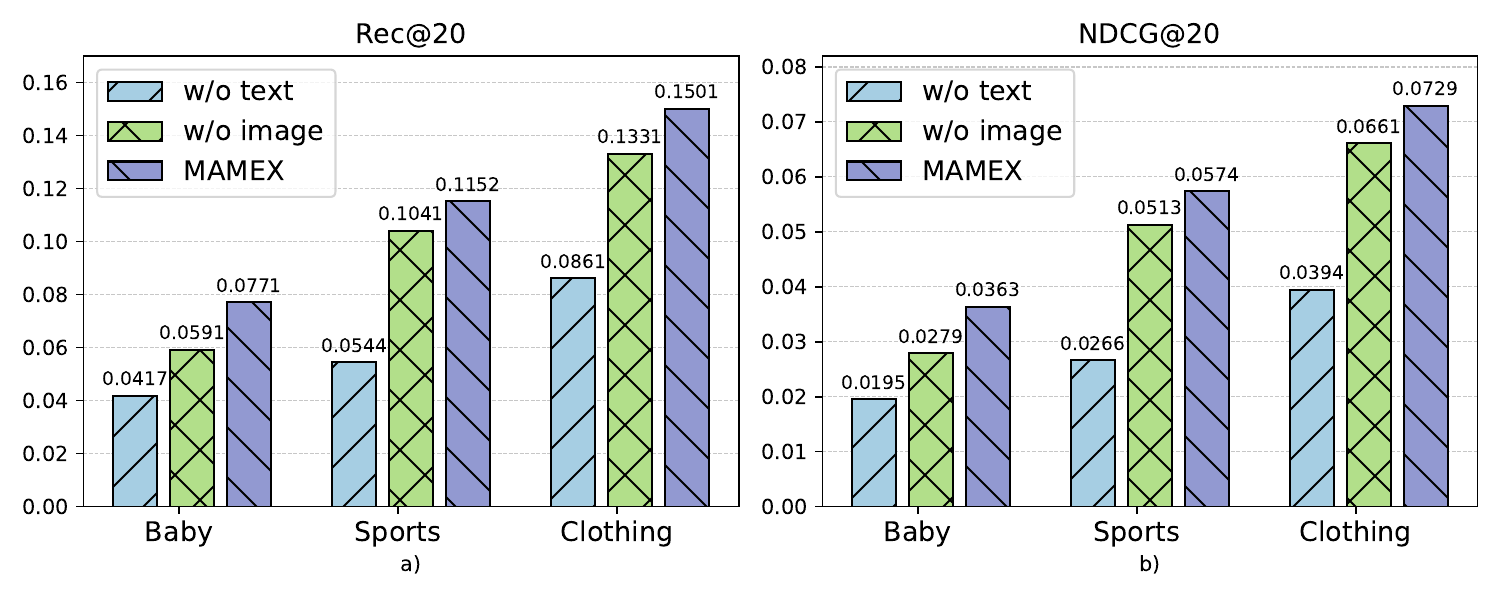}}
    \vspace*{-4mm}
    \caption{The impact of different modalities on three datasets.}
    \label{fig:modality}
\end{figure}

Figure~\ref{fig:modality} shows that the textual modality performs better than the visual modality, probably due to its richer semantics and more specific details, while images struggle with abstract attributes such as style or comfort. Furthermore, multimodal information surpasses any single modality. This finding underscores that the fusion approach successfully integrates multiple modalities, thus improving the overall performance of the MAMEX model.

\subsubsection{MoE Adapter Design Analysis}

\begin{figure}[ht]
  \includegraphics[width=0.47\textwidth]{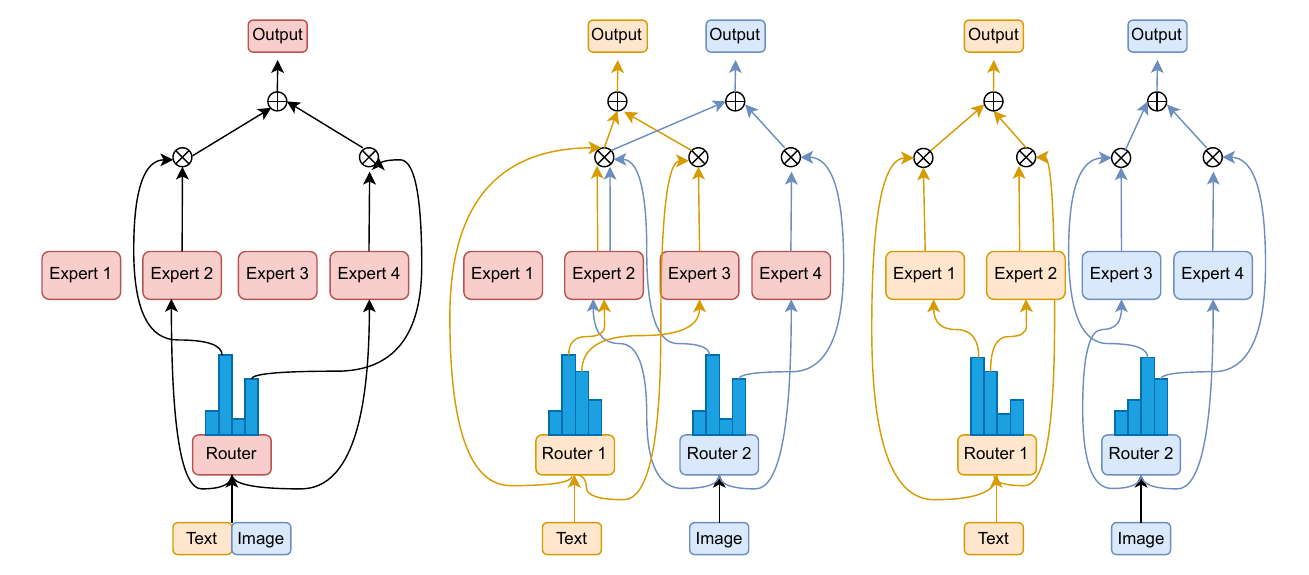}
  \caption{Three MoE adapter designs evaluated in our study.}
  \label{fig:moe-architectures}
\end{figure}

To further investigate fusion strategies, we evaluate three Mixture of Experts approaches: (a) \textbf{Joint Router}: concatenates all input modalities and employs a single router with shared experts; (b) \textbf{Modality-Specific Router}: uses separate routers for each modality while maintaining a common set of experts; and (c) \textbf{MAMEX}: assigns both dedicated routers and experts to each modality, thus facilitating maximum specialization, as illustrated in Figure ~\ref{fig:moe-architectures}.
MAMEX consistently outperforms both modality-specific router and joint router baselines in all evaluation metrics and datasets. On the Clothing dataset, it achieves relative gains of 0.94\% in Recall@20 and 1.72\% in NDCG@20 over the strongest baseline. Similarly, on the Sports dataset, it yields improvements of 3.61\% in Recall@20 and 5.60\% in NDCG@20. These consistent gains highlight the effectiveness of MAMEX’s structure, which combines modality-sensitive input with expert selection and specialized routing. This design enables more precise modeling of modality-specific features, surpassing the performance of conventional and partially specialized methods.

\begin{table}[ht]
  \centering
  \small
  \caption{Comparison of MoE Variants on Amazon Datasets.}
  \label{tab:moe-comparison}
  \begin{tabular}{l|cc|cc}
    \hline
    \textbf{Datasets} 
      & \multicolumn{2}{c|}{\textbf{Clothing}} 
      & \multicolumn{2}{c}{\textbf{Sports}} \\
    \hline
    \textbf{Metrics} 
      & \textbf{Rec@20} & \textbf{NDCG@20} 
      & \textbf{Rec@20} & \textbf{NDCG@20} \\
    \hline
    Joint Router   & 0.0847 & 0.2206 & 0.0664 & 0.1714 \\
    Mod-Specific   & 0.0847 & 0.2207 & 0.0640 & 0.1714 \\
    \textbf{MAMEX} & \textbf{0.0855} & \textbf{0.2245} 
                   & \textbf{0.0688} & \textbf{0.1810} \\
    \hline
  \end{tabular}
\end{table}

\section{Conclusion} 
This paper presents \textbf{MAMEX}, a dual-level Mixture of Experts framework for cold-start recommendation. By integrating modality-specific MoE layers with a learnable gating fusion, MAMEX captures modality-specific representations while dynamically balancing their contributions. Experiments on three Amazon datasets show consistent improvements in Recall and NDCG compared to state-of-the-art baselines, highlighting the effectiveness of the proposed architecture and regularization strategies. The core concept of MAMEX can also inspire research in multi-objective recommendation and interpretability. 
Future work should aim at addressing missing modalities, improving cross-modal generation, adding temporal MoE layers, and optimizing expert routing to enhance scalability and adaptability to evolving user preferences.

\bibliographystyle{ACM-Reference-Format}
\bibliography{sample-base}

\end{document}